\documentclass
[aps,prl,singlecolumn,preprint,showpacs,preprintnumbers,amsmath,amssymb,superscriptaddress]{revtex4}%
\usepackage{graphicx}
\usepackage{dcolumn}
\usepackage{bm}
\usepackage{amsmath}
\usepackage{amsfonts}
\usepackage{amssymb}%
\begin{document}
\preprint{APS/123-QED}
\title{Intrinsic and extrinsic origins of low-frequency noise in GaAs/AlGaAs
Schottky-gated nanostructures }
\author{Kenichi Hitachi}
\affiliation{NTT Basic Research Laboratories, NTT Corporation, Atsugi-shi, Kanagawa
243-0198, Japan}
\author{Takeshi Ota}
\affiliation{NTT Basic Research Laboratories, NTT Corporation, Atsugi-shi, Kanagawa
243-0198, Japan}
\author{Koji Muraki}
\affiliation{NTT Basic Research Laboratories, NTT Corporation, Atsugi-shi, Kanagawa
243-0198, Japan}
\date{\today}

\setlength{\baselineskip}{26pt}
\begin{abstract}
We study low-frequency noise in current passing through quantum point contacts
fabricated from several GaAs/AlGaAs heterostructures with different layer
structures and fabrication processes. In contrast to previous reports, there
is no gate-dependent random telegraph noise (RTN) originating from tunneling
through a Schottky barrier in devices fabricated using the standard low-damage
process. Gate-dependent RTN appears only in devices fabricated with a
high-damage process that induces charge trap sites. We show that the insertion
of AlAs/GaAs superlattices in the AlGaAs barrier helps to suppress trap
formation. Our results enable the fabrication of damage-resistant and thus
low-noise devices.

\end{abstract}

\maketitle

Low-frequency noise in the source-drain resistance of
metal-oxide-semiconductor and GaAs high-electron-mobility transistors arising
from background charge fluctuations have been a long-standing issue [1-3]. As
the device size decreases, device properties become increasingly susceptible
to local charge fluctuations. The noise can exhibit either $1/f$ or random
telegraphic behavior depending on the number of charge trap sites involved
[4]. In mesoscopic or nanoscale devices such as quantum point contacts (QPCs)
and quantum dots (QDs), single electron hopping occurring at somewhat distant
sites can be resolved as a measurable change in the device characteristics.
Because of their high charge sensitivity, QPCs have been widely used as charge
sensors for detecting individual charges in QDs, which constitutes a key
element of the present quantum information technology based on QDs. However,
the precision of the state readout is limited by the conductance fluctuation
of the QPC arising from uncontrolled background charge motion. Background
charge noise also causes fluctuations of the energy levels in QDs, which
results in decoherence and gate error in both charge [5-7] and spin qubits [8-10].

Several mechanisms have been considered as the origin of the low-frequency
noise in GaAs-based devices, including electron hopping between the
two-dimensional electron gas (2DEG) and trap sites [11, 12], current leakage
through the Schottky barrier [13-16], and electron hopping within the
remote-doping layer [17, 18]. An important observation in these studies is
that the low-frequency charge noise increases as the voltage $V_{g}$ applied
to the surface Schottky gate is decreased and becomes more negative. Since the
Schottky barrier becomes less opaque with decreasing $V_{g}$, the observed
$V_{g}$ dependence was interpreted as evidence that the current leakage
through the Schottky barrier was responsible for the charge noise. The strong
$V_{g}$ dependence of the noise, in turn, allowed the reduction of the charge
noise by employing bias cooling [14] or an additional global top gate [15],
both of which shift the operation point to a less negative voltage. In an
attempt to eliminate the charge noise associated with doping, single-electron
transistors based on undoped structures have been fabricated [19]; however,
the anticipated noise reduction has not yet been demonstrated.

In this paper, we study the effects of the layer structure and fabrication
process on the low-frequency current noise in QPC devices fabricated from
GaAs/Al$_{0.3}$Ga$_{0.7}$As modulation-doped heterostructures at 4.2 K to
clarify the origin of the noise. Specifically, in addition to the standard
delta-doped structure, we examined a structure in which AlAs/GaAs superlattice
(SL) barriers are inserted above and below the delta-doping layer---if the
current leakage through the Schottky barrier is the prime contributor to the
charge noise, the noise should be suppressed by employing the SL barriers, as
suggested in Ref. 15. As opposed to previous reports [14, 15] and the above
expectation, we found the noise in our devices to be nearly $V_{g}$
independent and unaffected by the SL barriers. On the other hand, devices
fabricated using a high-damage process that induces charge trap sites
exhibited strongly $V_{g}$-dependent noise characterized by random telegraphic
behavior. We show that the inclusion of SL barriers helps to suppress the
formation of trap sites and, accordingly, reduce the process-induced $V_{g}%
$-dependent noise. Our results pave the way to obtaining stable gated
nanostructures without bias cooling or a global top gate and assist in the
fabrication of damage-resistant devices.

We examined the use of devices with QPC and double quantum dot (DQD) gate
layouts, as shown in Fig. 1(a) and (b), respectively, to measure the charge
noise. A QPC device is defined by applying the same gate voltage $V_{g}$ to
the center gate C and one of the finger gates Q$_{i}$ ($i=1$-$4$). The gap
between C and Q$_{i}$ ranges from $W=190$ to $280$~nm (in 30-nm steps), which
allows the in-situ comparison of QPCs with different operation gate voltages.
For the DQD-type devices, a QPC was formed by applying the same $V_{g}$ to two
of eight gates, where the operation voltage was varied by activating a
different gate pair. We examined three kinds of layer structures with
conduction band profiles as schematically shown in Fig.~1(c)-(e). Type-I is a
standard delta-doped structure [Fig.~1(c)]. In type-I\hspace{-0.1em}I,
10-nm-thick AlAs/GaAs (2.1 nm/0.56 nm) SL barriers are inserted above and
below the delta-doping layer [Fig.~1(d)]. A uniform-doped structure
(type-I\hspace{-0.1em}I\hspace{-0.1em}I) was also examined [Fig.~1(e)].

The samples we investigated and the wafers we used are listed in Table I
together with their properties. In addition to the standard process that we
use to fabricate QPC and DQD devices, we also examined the effects of a
high-damage process by additionally employing intense UV ozone cleaning at
$100\ {}^{\circ}$C for 3 minutes prior to the electron-beam lithography. While
this eliminates the remaining photoresist and facilitates the lift-off
process, it also induces considerable damage, as evidenced by the decrease in
the carrier density and the mobility (sample E, Table I). For the structure
with the SL barriers, the same process had much less effect on the carrier
density and the mobility (sample F, Table I), which we discuss later. All
measurements were performed at 4.2 K. Unless otherwise noted, the results
presented below were obtained from devices fabricated with the standard process.

Figure~2(a) shows a typical measurement for sample A. We applied a
source-drain voltage of 300 $\mu$V and measured the current $I(t)$ through the
QPC over 20 seconds at several $V_{g}$ values with the QPC conductance
$G_{\mathrm{QPC}}$ below the first plateau, i.e., $0<G_{\mathrm{QPC}}%
<2e^{2}/h$, where $e$ is an elementary charge and $h$ is Planck's constant.
Following the method in Ref.~[20], we characterize the charge noise in terms
of the equivalent gate voltage noise defined as $\Delta V_{g}=\Delta
I/(d\bar{I}/dV_{g})$, where $\bar{I}$ and $\Delta I$ are the time average and
fluctuation of $I(t)$. Here, $\Delta I$ [Fig.~2(d)] was calculated as
\[
\Delta I=\sqrt{2 \int_{0.1}^{25}S_{I}(f)df}%
\]
using the Fourier power spectrum $S_{I}(f)$ of $I(t)$, and $\bar{I}$ and
$d\bar{I}/dV_{g}$ [Fig.~2(b) and (c)] were calculated numerically from the
data [21]. As shown in Fig. 2(d), $\Delta I$ clearly depends on $d\bar{I}/dV_{g}$,
which precludes the noise from the current amplifier as the origin of the
observed noise. We note that the voltage noise on the gates is also
negligible, as the resolution ($3$ $\mu V$) of the voltage sources that we
employed is about one order of magnitude smaller than the measured $\Delta
V_{g}$ [Fig.~2(e)]. As shown in Fig.~2(e), $\Delta V_{g}$ is not greatly
influenced by a change in $d\bar{I}/dV_{g}$ with $V_{g}$, and remains nearly
constant over the range $0<G_{\mathrm{QPC}}<2e^{2}/h$.

Figure 3 shows $\Delta V_{g}$ measured for QPCs fabricated from four different
wafers as a function of operating point $V_{g}$. Each data point corresponds to a different gate combination, representing an average taken over several $V_{g}$ values near the corresponding central operation voltage. The error bars indicate the standard deviation. We first note that, in the standard delta-doped structure (type-I), the measured noise is only weakly dependent on $V_{g}$. This also indicates that the measured noise level appropriately reflects the property of the sample and does not depend on the position or specific configuration of the QPC. Such $V_{g}$-independent noise is in sharp contrast to that described in Ref.~15, where $\Delta V_{g}$ increased exponentially as $V_{g}$ was decreased
below $-0.5$~V. As demonstrated in Refs.~14 and 15, $\Delta V_{g}$ increases
because decreasing $V_{g}$ makes the Schottky barrier less opaque and thus
enhances the tunneling probability from the gate to the trap sites in the
AlGaAs barrier. The absence of such strong $V_{g}$ dependence in our devices
suggests that tunneling through the Schottky barrier is irrelevant to the
observed noise. This conjecture is also supported by a comparison of the noise
levels of our devices and those studied in Ref.~15. Whereas at a small
negative $V_{g}$ the noise levels in samples A and B are twice that reported
in Ref.~15, due to the absence of $V_{g}$ dependence, these devices exhibit
much less noise at a large negative $V_{g}$. This enables us to define stable
nanostructures such as a QPC or a QD without employing bias cooling or an
additional global gate; this is a strong advantage, because it lifts the
restrictions on the operation voltage set by bias cooling or a global gate,
thereby greatly enhancing the degree of freedom in tailoring desired nanostructures.

As shown in Fig.~3, the insertion of the SL barriers did not reduce the noise
level in our devices (type-I\hspace{-0.1em}I). The independence of the charge
noise as regards the presence or absence of the SL barriers lends support to
the above conjecture that tunneling through the Schottky barrier is irrelevant
to the noise observed in our devices. The behavior of the sample fabricated
from the uniform-doped structure (type-I\hspace{-0.1em}I\hspace{-0.1em}I) is
similar to that of the other samples.

Another important feature of our data is that it is free from random
telegraphic, or switching behavior, even for large negative $V_{g}$ values, as
seen in Fig.~2(a). Indeed, we observed no switching noise, or random
telegraphic noise (RTN) for measurements obtained over a number of hours. This
was true for all the devices that we examined that were fabricated with the
standard process, independent of the type of layer structure or doping. Note
that RTN appears when only a few trap sites located near the QPC are the prime
contributors to the charge fluctuation. The absence of RTN in our devices,
combined with the low noise level at a large negative $V_{g}$, implies that in
our devices there is no trap site available for tunneling electrons near the
QPC. This also suggests that switching noise due to tunneling through a
Schottky barrier is not necessarily inherent to GaAs-based modulation-doped structures.

The absence of RTN in our devices in turn suggests the existence of extrinsic
origins of RTN. To explore this possibility, we examined the effects of
process-induced charge traps by additionally employing intense UV ozone
cleaning prior to the electron-beam lithography. This process induces charge
trap sites, as evidenced by the significant decrease in the mobility (sample
E, Table I). As shown in the inset to Fig. 4(a), we did observe RTN in the
device fabricated with the intense UV cleaning. In Fig.~4(a), we plot $\Delta
V_{g}$ of this device as a function of $V_{g}$ and compare it with the results
obtained for a device fabricated from the same wafer using the standard
process. The intense UV\ cleaning induced strongly $V_{g}$-dependent noise
that increased rapidly below $-0.5$ V, which confirms the link between strong
$V_{g}$ dependence and random telegraphic behavior.

Interestingly, we found that the SL barriers played an unexpected role that
would allow us to make devices much more resistant to damage from processing.
As shown in Table I, the inclusion of SL\ barriers (sample F) attenuates the
deleterious effect that processing has on carrier density and mobility. This
is particularly clear for the mobility, which differs by almost one order of
magnitude. This result clearly shows that the SL barriers help to suppress the
formation of trap sites. The reduced trap density is also evident in the noise
behavior. In contrast to the results for the samples without SL barriers,
$\Delta V_{g}$ is barely affected by the high-damage process and remains
almost $V_{g}$ independent as shown in Fig.~4(b). It is possible that the SL
barriers block tunneling through the Schottky barrier and thereby suppress the
$V_{g}$-dependent RTN. In the present case, however, it is more likely that
the reduced RTN is due to the lower trap density, as clearly demonstrated by
the electron density and the mobility (Table~I). Although the exact mechanism
for the suppressed trap formation is unknown, our finding enables the
fabrication of damage-resistant and thus low-noise devices. 

Finally, we discuss the origin of the $V_{g}$-independent noise observed in
the samples fabricated with the standard low-damage process. In the samples
fabricated with the high-damage process, the spatial range over which the trap
sites relevant to the RTN are distributed can be crudely estimated from the
size of the current step at each switching event. In our DQD-type devices, the
operating voltage of the QPC charge sensor shifts by 1.3 (0.9) mV (data not
shown) when one electron is added to or removed from the QD on the near (far)
side about 300 (400) nm from the QPC. Since the observed RTN corresponds to a
shift in the QPC operating voltage a few times larger than these values, we
can estimate that the relevant trap sites are distributed within $\sim100$~nm
of the QPC. This estimate, which also implies that the relevant trap sites are
located close to the split gate, is consistent with the noise being strongly
$V_{g}$ dependent. In turn, it follows that the $V_{g}$-independent noise
dominant in our devices fabricated with the standard process is due to charge
hopping taking place at distant sites distributed over a much wider spatial
range. This picture is consistent with the absence of switching behavior and
$V_{g}$ dependence. As we discussed above, vertical tunneling through the
AlGaAs (or SL)\ barrier is unlikely to be its origin. We therefore speculate
that charge hopping within the remote-doping layer is the most likely cause of
the $V_{g}$-independent noise. However, we do not know why the noise level differs among wafers with the identical doping (compare samples A, B, and C in Fig. 3, which all have the same doping level). Indeed, even though device E has high charge trap density, it shows lower noise than sample F at $V_{g} = -0.4$ V, where RTN is absent (Fig. 4). This result may be attributed to the difference in the starting material. If so, it implies that these process-induced traps do not contribute to the $V_{g}$-independent noise. A further investigation is necessary to clarify the origin of the $V_{g}$-independent noise.

In summary, we identified two types of low-frequency noise in GaAs/AlGaAs QPCs
distinguished by $V_{g}$ dependence linked with the presence/absence of
switching behavior. Our results indicate that the commonly observed $V_{g}%
$-dependent switching noise is not necessarily an intrinsic property of
modulation-doped GaAs/AlGaAs heterostructures and point to the importance of
eliminating extrinsic origins of low-frequency noise. We found that the
inclusion of SL barriers attenuates damage from processing that can be a
source of low-frequency noise. Our findings will thus enhance the degree of
freedom in tailoring low-noise Schottky-gated nanostructures and assist in the
fabrication of devices that are resistant to damage from processing.

Part of this work was supported financially by the MEXT Grant-in-Aid for
Scientific Research on Innovative Areas (21102003), and the Funding Program
for World-Leading Innovative R\&D on Science and Technology (FIRST).

\bigskip

[1] M. J. Kirton and M. J. Uren, Advances in Physics, \textbf{38}, 367 (1989).

[2] Y. P. Li, D. C. Tsui, J. J. Heremans, and J. A. Simmons, Appl. Phys. Lett.
\textbf{57}, 774 (1990).

[3] D. H. Cobden, N. K. Patel, M. Pepper, D. A. Ritchie, J. E. F. Frost, and
G. A. C. Jones, Phys. Rev. B \textbf{44}, 1938 (1991).

[4] S. Machlup, J. Appl. Phys. \textbf{25}, 341 (1954).

[5] T. Hayashi, T. Fujisawa, H. D. Cheong, Y. H. Jeong, and Y. Hirayama, Phys.
Rev. Lett. \textbf{91}, 226804 (2003).

[6] K. D. Petersson, J. R. Petta, H. Liu, and A. C. Gossard, Phys. Rev. Lett.
\textbf{105}, 246804 (2010).

[7] T. Itakura and Y. Tokura, Phys. Rev. B \textbf{67}, 195320 (2003).

[8] J. R. Petta, A. C. Johnson, J. M. Taylor, E. A. Laird, A. Yacoby, M. D.
Lukin, C. M. Marcus, M. P. Hanson, and A. C. Gossard, Science \textbf{309},
2180 (2005).

[9] Q. Li, L. Cywinski, D. Culcer, X. Hu, and S. Das Sarma, Phys. Rev. B
\textbf{81}, 085313 (2010).

[10] D. Culcer, X. Hu, and S. Das Sarma, Appl. Phys. Lett. \textbf{95}, 073102 (2009).

[11] C. Dekker, A. J. Scholten, F. Liefrink, R. Eppenga, H. van Houten, and C.
T. Foxon, Phys. Rev. Lett. \textbf{66}, 2148 (1991).

[12] T. Sakamoto, Y. Nakamura, and K. Nakamura, Appl. Phys. Lett. \textbf{67},
2220 (1995).

[13] D. H. Cobden, A. Savchenko, M. Pepper, N. K. Patel, D. A. Ritchie, J. E.
F. Frost, and G. A. C. Jones, Phys. Rev. Lett. \textbf{69}, 502 (1992).

[14] M. Pioro-Ladriere, J. H. Davies, A. R. Long, A. S. Sachrajda, L.
Gaudreau, P. Zawadzki, J. Lapointe, J. Gupta, Z. Wasilewski, and S.
Studenikin, Phys. Rev. B \textbf{72}, 115331 (2005).

[15] C. Buizert, F. H. L. Koppens, M. Pioro-Ladriere, H-P. Tranitz, I. T.
Vink, S. Tarucha, W. Wegscheider, and L. M. K. Vandersypen, Phys. Rev. Lett.
\textbf{101}, 226603 (2008).

[16] Y. X. Liang, Q. Dong, M. C. Cheng, U. Gennser, A. Cavanna, and Y. Jin,
Appl. Phys. Lett. \textbf{99}, 113505 (2011).

[17] G. Timp, R. E. Behringer, and J. E. Cunningham, Phys. Rev. B \textbf{42},
9259 (1990).

[18] C. Kurdak, C. J. Chen, D. C. Tsui, S. Parihar, S. Lyon, and G. W.
Weimann, Phys. Rev. B \textbf{56}, 9813 (1997).

[19] A. M. See, O. Klochan, A. R. Hamilton, A. P. Micolich, M. Aagesen, and P.
E. Lindelof, Appl. Phys. Lett. \textbf{96}, 112104 (2010).

[20] S. W. Jung, T. Fujisawa, Y. Hirayama, and Y. H. Jeong, Appl. Phys. Lett.
\textbf{85}, 768 (2004).

[21] We have confirmed that the high-frequency components above 25 Hz have a negligible contribution to $\Delta I$.

\bibliographystyle{plain}
\bibliography{basename of .bib file}

\begin{figure}
\includegraphics[width=\linewidth]{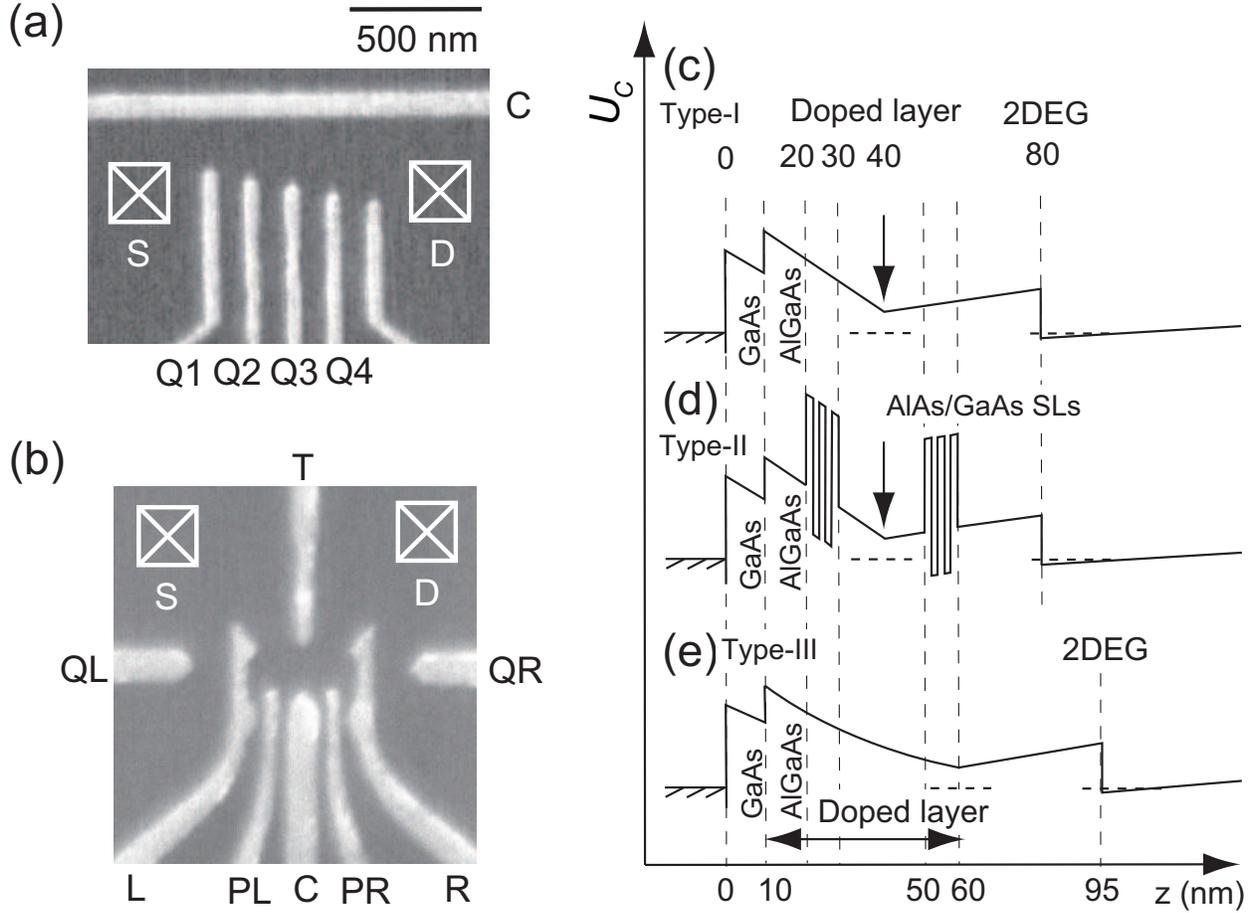}
\caption{(Color online) ~(a), (b) SEM images showing the gate layouts for (a) QPC and (b) DQD devices. (c)-(e) Schematic illustrations of conduction band profiles for (c) a standard delta-doped structure (type-I), (d) a delta-doped structure with AlAs/GaAs SL barriers (type-II), and (e) an uniform-doped structure (type-III). The Si doping density is $4\times10^{12}$ cm$^{-2}$ for delta doping and $1\times10^{18}$ cm$^{-3}$ for uniform doping. }
\end{figure}
\begin{figure}
\includegraphics[width=\linewidth]{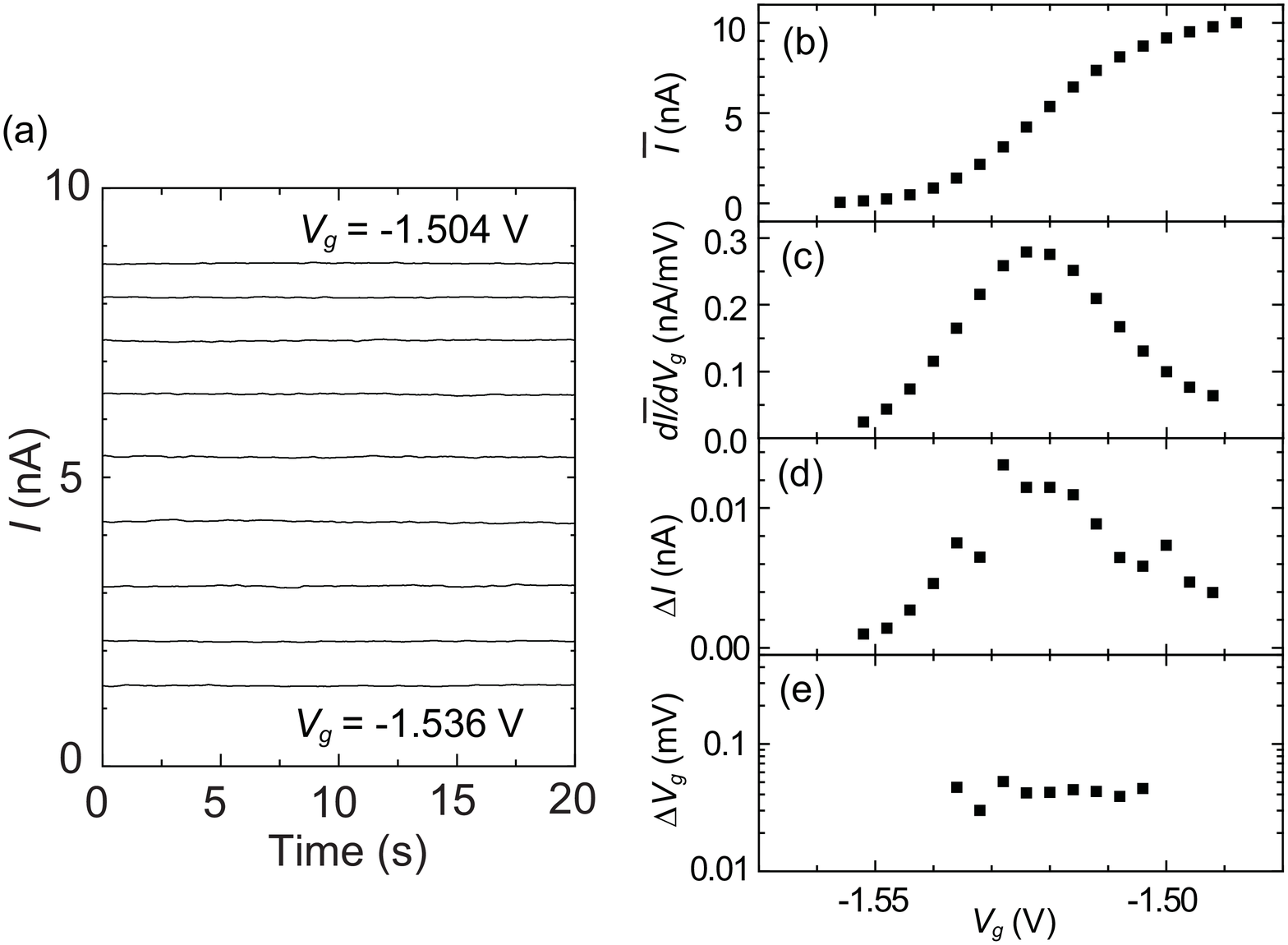}
\caption{(Color online) ~(a) Time traces of current $I(t)$ through the QPC in sample A\ (type-I) for several $V_{g}$ values. (b)-(e) Analysis of the time traces $I(t)$ shown in (a): (b) average current $\overline{I}$, (c) transconductance $\mathrm{d}\overline{I}/\mathrm{d}V_{g}$, (d) current fluctuation $\Delta I$, and (e) equivalent gate voltage noise $\Delta V_{g}$, plotted as a function of $V_{g}$.}
\end{figure}
\begin{figure}
\includegraphics[width=0.7\linewidth]{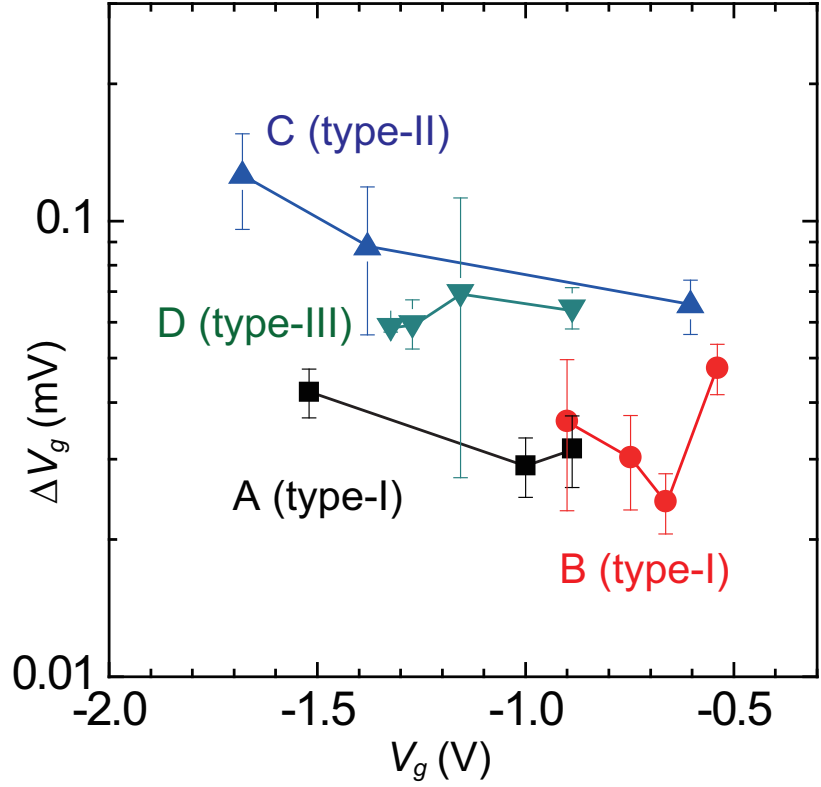}
\caption{(Color online) Equivalent gate voltage noise $\Delta V_{g}$ as a function of $V_{g}$ for samples A to D fabricated from different wafers with various layer structures and doping schemes (see Table I).}
\end{figure}
\begin{figure}
\includegraphics[width=0.7\linewidth]{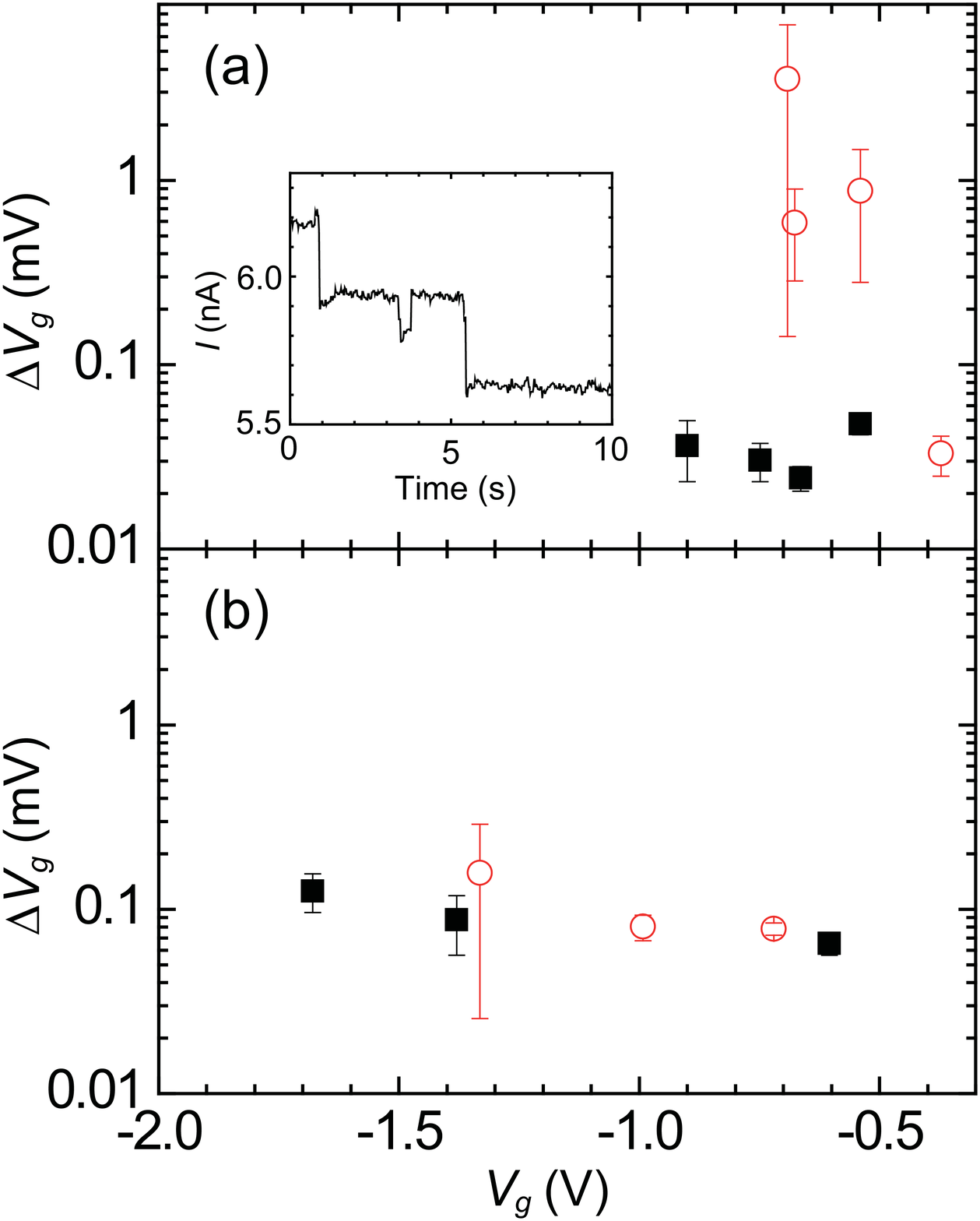}
\caption{(Color online) Comparison of equivalent gate voltage noise $\Delta V_{g}$ in samples fabricated from the same wafer using the standard process (solid squares) and with additional intense UV cleaning (open circles): (a) standard delta-doped structure (type-I: samples B and E) and (b) delta-doped structure with SL barriers (type-II: samples C and F). Inset to (a): typical time trace of QPC current in a sample fabricated with intense UV cleaning (sample E).}
\end{figure}
\begin{table}
\caption{List of samples and wafers used. The carrier density $n$ and the mobility $\mu$ are shown in units of $10^{11}$~cm$^{-2}$ and $10^{6}$~cm$^{2}$/Vs, respectively. \label{}}
\begin{tabular}[b]{|c|c|c|c|c|c|c|}
\hline \hline
sample & \shortstack{wafer \\ \#} & \shortstack{wafer \\ structure} & \shortstack{fabrication \\ process} & \shortstack{$n$} & \shortstack{$\mu$} & \shortstack{gate \\layout} \\ \hline
A & \ R193  \ & \ type-I \  & \ standard \ & \ 2.3 \  & \ 1.8 \ & \ DQD \ \\
B & \ R244  \ & \ type-I \  & \ standard \ & \ 2.2 \  & \ 1.3 \ & \ QPC \ \\
C & \ R219  \ & \ type-I\hspace{-.1em}I \  & \ standard \ & \ 2.3 \  & \ 1.3 \  & \ DQD \ \\
D & \ R247  \ & \ type-I\hspace{-.1em}I\hspace{-.1em}I \  & \ standard \ & \ 2.5 \  & \ 2.9 \ & \ QPC \ \\
E & \ R244  \ & \ type-I   & \ high-damage \  & \ 1.9 \ & \ 0.14 \ & \ QPC \ \\
F & \ R219  \ & \ type-I\hspace{-.1em}I \ & \ high-damage \ & \ 2.3 \ & \ 1.2 \  & \ DQD \ \\
\hline
\end{tabular}
\end{table}

\end{document}